\documentclass[a4,11pt]{article}

\usepackage{newlfont}
\usepackage[small]{caption2}
\usepackage[below]{placeins}
\usepackage{flafter}
\usepackage{amsfonts}
\usepackage{amsmath}
\usepackage{amssymb}
\usepackage{mathtools}
\usepackage[american]{babel}
\usepackage{graphicx}
\usepackage{subfigure}
\usepackage{multirow}
\usepackage{url}
\usepackage{tikz}
\usetikzlibrary{positioning,chains,fit,shapes,calc,decorations.markings}
\usepackage{calc}
\tikzstyle{vertex}=[circle, draw, inner sep=0pt, minimum size=6pt]
\tikzstyle{marked}=[circle,draw,dotted,inner sep=0pt, minimum size=6pt]

\allowdisplaybreaks

\newcommand{\ket}[1]{\big| #1 \big\rangle}
\newcommand{\bra}[1]{\big\langle #1 \big|}
\newcommand{\bracket}[3]{\big\langle #1 \big| #2 \big| #3 \big\rangle}       

\begin{document}


\title{Szegedy's quantum walk with queries}
\author{Raqueline A. M. Santos}
\date{
{\small Faculty of Computing - University of Latvia}
\\{\small Raina bulv. 19, Riga, LV-1586, Latvia}
\\{\small \url{rsantos@lu.lv}}
}

\maketitle

\begin{abstract}
When searching for a marked vertex in a graph, Szegedy's usual search operator is defined by using the transition probability matrix of the random walk with absorbing barriers at the marked vertices. Instead of using this operator, we analyze searching with Szegedy's quantum walk by using reflections around the marked vertices, that is, the standard form of quantum query. We show we can boost the probability to 1 of finding a marked vertex in the complete graph. Numerical simulations suggests that the success probability can be improved for other graphs, like the two-dimensional grid. We also prove that, for a certain class of graphs, we can express Szegedy's search operator, obtained from the absorbing walk, using the standard query model.
\end{abstract}

\section{Introduction}

Since the introduction of the first quantization of a random walk, quantum walks have been used in the development of quantum algorithms that outperform their classical versions~\cite{Shenvi:2003,Childs:2004,Ambainis:2004,Ambainis:2005,Krovi:2010,Portugal:2013}. 
The first model was the coined discrete time quantum walk on the line by Aharonov~{\it et al.}~\cite{Aharonov:1993}, followed by the continuos time quantum walk model by Farhi~{\it et al.}~\cite{Farhi:1998}.
In 2004, Szegedy~\cite{Szegedy:2004} proposed a coinless quantum walk model which is driven by reflection operators in a bipartite graph. The evolution operator depends directly on the transition probability matrix of the Markov chain associated to the graph.
Szegedy showed that the problem of detecting can be solved quadratically faster than classically for ergodic and symmetric Markov chains. The problem of spatial searching was also analyzed in this model by Refs.~\cite{Magniez:2011,Krovi:2010}.
Santos and Portugal~\cite{Santos:2010} analyzed the analytical details of the search on the complete graph.



For searching with quantum walks we need to find a way to differentiate between marked and non-marked vertices. Szegedy described in his model an evolution operator which acts differently on the marked vertices by using the transition probability matrix of the absorbing walk on the graph. In this way, Szegedy was able to define a quantum hitting time which is analogous to the classical definition. Searching with the coined quantum walk model is usually done by applying a different coin operator which flips the phase of the marked vertices~\cite{Shenvi:2003,Ambainis:2005}. Inspired by that, we analyze what happens in Szegedy's model when we add a reflection that flips the phase of the marked vertices, that is, the standard form of quantum query. We show that the success probability of finding a marked vertex in the complete graph boosts from $1/2$ to $1$. The analysis is also done for multiple marked vertices. Numerical simulations suggests that Szegedy's quantum walk with queries can also improve the success probability for other graphs. We use the two-dimensional lattice as an example. 
Different operators, with different combinations of reflections around the marked vertices, are also analyzed. 
By identifying a similar behavior between two of these operators, we prove that for any strongly regular graph with one marked vertex, we can write Szegedy's search operator, obtained from the absorbing walk, by using the standard query model.


Recently, Portugal~{\it et al.}~\cite{Portugal:2015} showed the connection between Szegedy's model and the staggered quantum walk model. They showed that Szegedy's quantum walk can be cast into the staggered model, including its search operator. 
We show that the search operators that will be here described using Szegedy's model can still be cast in the staggered model.
Later Portugal~\cite{Portugal:2016} also showed the connection between the coined discrete-time quantum walk model and Szegedy's model. The standard flip-flop quantum walks with Hadamard or Grover coin can be translated into Szegedy's quantum walk on a bipartite graph $(X,Y)$, where the set of vertices $Y$ have degree two and the weights associated with the edges incident on those vertices must be equal. In this paper, we deal with the case which the equivalence is not yet known. 

The paper is organized as follows. In Sec.~\ref{sec:qw}, we review Szegedy's quantum walk and its way of doing search. The results for the complete graph~\cite{Santos:2010} are summarized. In Sec.~\ref{sec:qww}, we introduce the Szegedy's quantum walk with queries and we analyze its behavior on the complete graph for one marked vertex and multiple marked vertices. In Sec.~\ref{sec:oso}, we numerically analyze the behavior of other operators and we show that Szegedy's operator can be written using queries for some class of graphs. The conclusions and discussions are drawn in Sec.~\ref{sec:conc}.

\section{Szegedy's Quantum Walk}\label{sec:qw}
Let $\Gamma(X,E)$ be a connected and undirected graph, where $X$ is the set of vertices and $E$ is the set of edges. Consider that $P$ is the stochastic matrix associated to the graph and $p_{xy}$ are its components. Szegedy~\cite{Szegedy:2004} has proposed a quantum walk driven by reflection operators. We associate with the graph a Hilbert space ${\cal H}^{n^2} = {\cal H}^{n}\otimes {\cal H}^{n} $, where $ n = | X |$. The computational basis of $ {\cal H}^{n^2} $ is $ \big \{\ket {x, y}: x \in X, y \in X \big \} $. The evolution operator $ U_P $ is given by
\begin{equation}\label{ht_U_ev}
    U_P := {\cal R}_B \, {\cal R}_A,
\end{equation}
where
\begin{eqnarray}
  {\cal R}_A &=& 2\sum_{x\in X}\ket{\Phi_x}\bra{\Phi_x} - I_{n^2}, \label{ht_RA}\\
  {\cal R}_B &=& 2\sum_{y\in X}\ket{\Psi_y}\bra{\Psi_y} - I_{n^2}, \label{ht_RB}
\end{eqnarray}
are reflections around the subspaces generated by $\ket{\Phi_x}$ and $\ket{\Psi_y}$, respectively, and
\begin{eqnarray}
  \ket{\Phi_x} &=& \ket{x}\otimes \left(\sum_{y\in Y} \sqrt{p_{x y}} \, \ket{y}\right), \label{ht_alpha_x} \\
  \ket{\Psi_y}  &=&  \left(\sum_{x\in X} \sqrt{p_{y x}} \, \ket{x}\right)\otimes \ket{y}. \label{ht_beta_y}
\end{eqnarray}
We can see the states $\ket{\Phi_x}$ and $\ket{\Psi_y}$ as superpositions over the edges that start from vertex $x$ and vertex $y$, respectively.  

Szegedy's quantum walk can be interpreted as a quantum walk on the edges of the graph, where $\ket{x,y}$ is the edge which represents the walker being in vertex $x$ coming from vertex $y$. Or according to~\cite{Portugal:2015}, it can also be seen as a quantum walk on the linegraph of the bipartite graph obtained after the duplication process. Szegedy's quantum walk was originally designed for a bipartite graph, with two sets of vertices $X$ and $Y$. If we have a graph with only one set of vertices $X$, we can duplicate it, by doing $Y=X$ and making the connections between the two sets as in the original graph. See Fig.~\ref{fig1} for an example. So, the Hilbert space ${\cal H}^{n}\otimes {\cal H}^{n}$ is associated to the obtained bipartite graph or to the edges of the original graph, as you may prefer.

\subsection{Searching}\label{sec:sw}
When dealing with the problem of detecting or finding marked vertices, it is important to differentiate the behavior of the evolution operator on these special vertices.
 Instead of using the stochastic matrix $P$, Szegedy used a modified evolution operator $U_{P'}$ associated with a modified stochastic matrix. $P^\prime$ represents the classical absorbing random walk in the marked vertices and  is given by
\begin{equation}\label{ht_pprime}
    p_{x y}^\prime = \left\{
                       \begin{array}{ll}
                         p_{x y}, & \hbox{$x\not\in M$;} \\
                         \delta_{x y}, & \hbox{$x\in M$,}
                       \end{array}
                     \right.
\end{equation}
where $M$ is the set of marked vertices. The initial state of the quantum walk is a superposition over all edges of the graph,
\begin{equation}
    \ket{\psi(0)} = \frac{1}{\sqrt n} \sum_{
x,y\in X
} \sqrt{p_{xy}} \ket{x,y}.
\end{equation}
The operator $U_{P^\prime}$ acts differently on the marked vertices. It makes the probability on the marked vertices grow in the beginning of the evolution, but it will not make the walker stay put on it as in the classical case. As we are dealing with a unitary operator, the probability will oscillate through time.

\begin{figure}[!htb]
\centering
\subfigure[fig1][$n=5$]{
\centering
\begin{tikzpicture}[thick,shorten >=0pt,-]
  \tikzstyle{vertex}=[circle,draw=black,fill=black!15,minimum size=12pt,inner sep=0pt]

  \foreach \name/\angle/\text in {P-4/234/4, P-5/162/5, 
                                  P-1/90/1, P-2/18/2, P-3/-54/3}
    \node[vertex,xshift=6cm,yshift=0.5cm] (\name) at (\angle:1cm) {\footnotesize{${\text}$}};

  \foreach \from/\to in {1/2,2/3,3/4,4/5,5/1,1/3,2/4,3/5,4/1,5/2}
    { \draw (P-\from) -- (P-\to);}

\end{tikzpicture}
\hspace{0.5cm}
\begin{tikzpicture}[thick,every node/.style={draw,circle},-,shorten >= 0pt,shorten <= 0pt]

\begin{scope}[start chain=going below,node distance=4mm]
\foreach \i in {1,2,...,5}
  \node[fill=black!15,on chain,minimum size=12pt,inner sep=0pt] (f\i) {\footnotesize{$\i$}};
\end{scope}

\begin{scope}[xshift=2cm,yshift=0.0cm,start chain=going below,node distance=4mm]
\foreach \i in {1,2,...,5}
  \node[fill=black!15,on chain,minimum size=12pt,inner sep=0pt] (s\i) {\footnotesize{$\i$}};
\end{scope}

\draw (f1) -- (s2);
\draw (f1) -- (s3);
\draw (f1) -- (s4);
\draw (f1) -- (s5);
\draw (f2) -- (s1);
\draw (f2) -- (s3);
\draw (f2) -- (s4);
\draw (f2) -- (s5);
\draw (f3) -- (s1);
\draw (f3) -- (s2);
\draw (f3) -- (s4);
\draw (f3) -- (s5);
\draw (f4) -- (s1);
\draw (f4) -- (s2);
\draw (f4) -- (s3);
\draw (f4) -- (s5);
\draw (f5) -- (s1);
\draw (f5) -- (s2);
\draw (f5) -- (s3);
\draw (f5) -- (s4);
\end{tikzpicture}
\label{fig1}
}\hspace{1cm}
\subfigure[fig1][$n=5$ and $m=1$]{
\centering
\begin{tikzpicture}[thick,shorten >=0pt,-]
  \tikzstyle{vertex}=[circle,draw=black,fill=black!15,minimum size=12pt,inner sep=0pt]
  \tikzstyle{vertexm}=[circle,draw=black,fill=red!15,minimum size=12pt,inner sep=0pt]
  \tikzset{every loop/.style={min distance=5mm,in=100,out=160,looseness=7}}
  
  \foreach \name/\angle/\text in {P-4/234/4, P-1/90/1, P-2/18/2, P-3/-54/3}
        \node[vertex,xshift=6cm,yshift=.5cm] (\name) at (\angle:1cm) {\footnotesize{$\text$}};
  
  \node[double,vertexm,xshift=6cm,yshift=.5cm] (P-5) at (162:1cm) {\footnotesize{$5$}};
  
  \foreach \from/\to in {1/2,2/3,3/4,1/3,2/4,3/5,4/1}
    { \draw (P-\from) -- (P-\to);}
  
  \foreach \from/\to in {1/5,4/5,3/5,2/5}
    { \draw[->] (P-\from) -- (P-\to);} 
    
   \draw[->]  (P-5) edge [loop below] (P-5);
   
\end{tikzpicture}
\hspace{0.5cm}
\begin{tikzpicture}[thick,every node/.style={draw,circle},fsnode/.style={fill=black!15},ssnode/.style={fill=black!15},every fit/.style={ellipse,draw,inner sep=-2pt,text width=2cm},-,shorten >= 0pt,shorten <= 0pt]

\begin{scope}[start chain=going below,node distance=4mm]
\foreach \i in {1,2,...,4}
  \node[fsnode,on chain,minimum size=12pt,inner sep=0pt] (f\i) {\footnotesize{$\i$}};
\node[double,fill=red!15,on chain,minimum size=12pt,inner sep=0pt] (f5) {\footnotesize{$5$}};
\end{scope}
  
\begin{scope}[xshift=2cm,yshift=0.0cm,start chain=going below,node distance=4mm]
\foreach \i in {1,2,...,4}
  \node[ssnode,on chain,minimum size=12pt,inner sep=0pt] (s\i) {\footnotesize{$\i$}};
\node[double,fill=red!15,on chain,minimum size=12pt,inner sep=0pt] (s5) {\footnotesize{$5$}};
\end{scope}

\draw (f1) -- (s2);
\draw (f1) -- (s3);
\draw (f1) -- (s4);
\draw[->] (f1) -- (s5);
\draw (f2) -- (s1);
\draw (f2) -- (s3);
\draw (f2) -- (s4);
\draw[->] (f2) -- (s5);
\draw (f3) -- (s1);
\draw (f3) -- (s2);
\draw (f3) -- (s4);
\draw[->] (f3) -- (s5);
\draw (f4) -- (s1);
\draw (f4) -- (s2);
\draw (f4) -- (s3);
\draw[->] (f4) -- (s5);
\draw[->] (s1) -- (f5);
\draw[->] (s2) -- (f5);
\draw[->] (s3) -- (f5);
\draw[->] (s4) -- (f5);
\draw (f5) -- (s5);
\end{tikzpicture}
}
\caption{Complete graphs with 5 vertices and their associated bipartite graphs. (a) The graphs associated to $P$.  (b) The graphs associated to $P'$ with one marked vertex.}
\label{fig:cgs}
\end{figure}
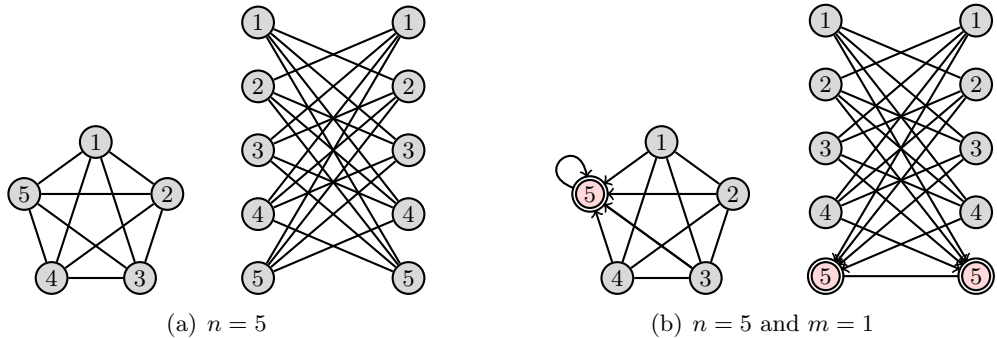

\begin{figure}[!htb]
\centering
\includegraphics[width=2.5in]{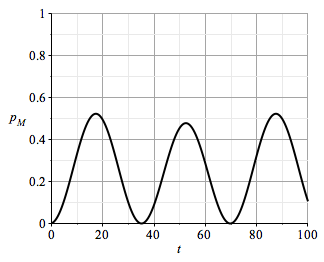}
\caption{Probability of finding a marked vertex on the complete graph with $n=1000$ and $m=1$. The maximum probability of $\approx 0.52$ is achieved at time $t = 17$.} \label{fig:pmtup}
\end{figure}

For example, consider the complete graph with $n$ vertices and $m = |M|$ marked vertices. In Fig.~{\ref{fig:cgs}} we can see the graphs associated to $P$ and $P'$.
The initial state for the complete graph is
\begin{equation}\label{ht_ini_cond_CG}
    \ket{\psi(0)} = \frac{1}{\sqrt{n(n-1)}} \sum^n_{\begin{subarray}{c}
x, y =1\\
 x\neq y
\end{subarray}}\ket{x}\ket{y}.
\end{equation}
Let
\begin{equation}
p_M(t) = \bracket{\psi(t)}{\sum_{x \in M}\ket{x}\bra{x}\otimes I_n}{\psi(t)}
\end{equation}
be the probability of finding a marked vertex, where $\ket{\psi(t)} = U_{P'}^t\ket{\psi(0)}$. According to~\cite{Santos:2010}, we can write
\begin{equation}
\begin{split}
  p_M(t)  &= \dfrac{m(m-1)}{n(n-1)}+\dfrac{m(n-m)}{n(n-1)}\left(\frac {n-1}{ 2\,n-m-2} T_{2t}\left(\frac{n-m-1}{n-1}\right) + \right.\\
& \left.U_{2t-1}\left(\frac{n-m-1}{n-1}\right)+ \dfrac{n-m-1}{2n-m-2}\right)^{2},
\end{split}
\label{ht_pM_CG}
\end{equation}
where $T_n$ and $U_n$ are the $n$-th Chebyshev polynomial of the first  and second kind, respectively.
The graph of $p_M(t)$ is depicted in Fig.~{\ref{fig:pmtup}} when $n=1000$ and $m=1$. 

The first point of maximum occurs at time 
\begin{equation}
t_{\textrm{max}}=
\dfrac{\arctan\left(\dfrac{\sqrt{2n-m-2}}{\sqrt{m}}\right)}{2\arccos\left(\dfrac{n-m-1}{n-1}\right)} = \dfrac{\pi}{4}\sqrt{\dfrac{n}{2\,m}}-\frac{1}{4}+O\left({\frac{1}{\sqrt n}}\right),
\label{eq:tmax}
\end{equation}
and by doing a measurement at this time give us the probability
\begin{equation}
p_{M}(t_{\textrm{max}}) = \dfrac{1}{2}+\sqrt{\dfrac{m}{2\,n}}+O\left({\frac{1}{n}}\right).
\end{equation}

\section{Szegedy's quantum walk with queries}\label{sec:qww}
An oracle is a black box which gives the algorithm information about the marked vertices or the solution of the problem we are trying to solve. 
The standard oracle query maps $\ket{x}$ to $-\ket{x}$, if $x \in M$, and $\ket{x}$ to $\ket{x}$, otherwise. It has been used for searching in the coined quantum walk model~\cite{Shenvi:2003,Ambainis:2005}. We can represent this oracle query as a reflection around the marked vertices, that is,
\begin{equation}
{\cal R}_M = I_n-2\sum_{x\in M}\ket{x}\bra{x}.
\end{equation} 
Szegedy's quantum walk uses a Hilbert space which has two registers. We will denote ${{\cal R}_M}_1 = {\cal R}_M\otimes I_n$ the reflection around the marked vertices acting in the first register.
Instead of using $U_{P'}$, let us use the following evolution operator
\begin{equation}
U_M=U_P{{\cal R}_M}_1 = {\cal R}_{\cal B} \, {\cal R}_{\cal A}{{\cal R}_M}_1
\end{equation}
which is simply Szegedy's quantum walk with queries. 
Next we will see how the operator $U_M$ behaves on the complete graph.

\subsection{Search on the complete graph}

\begin{figure}[!htb]
\centering
\begin{tikzpicture}[thick,shorten >=0pt,-]
  \tikzstyle{vertex}=[circle,draw=black,fill=black!15,minimum size=12pt,inner sep=0pt]
\tikzstyle{vertexm}=[double,circle,draw=black,fill=red!15,minimum size=12pt,inner sep=0pt]
  \foreach \name/\angle/\text in {P-4/234/a, 
                                  P-1/90/a, P-2/18/a, P-3/-54/a}
    \node[vertex,xshift=6cm,yshift=.5cm] (\name) at (\angle:1cm) {\footnotesize{$\text$}};

\node[vertexm,xshift=6cm,yshift=.5cm] (P-5) at (162:1cm) {\footnotesize{$b$}};
  \foreach \from/\to in {1/2,2/3,3/4,4/5,5/1,1/3,2/4,3/5,4/1,5/2}
    { \draw (P-\from) -- (P-\to);}

\end{tikzpicture}
\hspace{1cm}
\begin{tikzpicture}[thick,
  every node/.style={draw,circle},
  fsnode/.style={fill=black!15},
  ssnode/.style={fill=black!15},
  every fit/.style={ellipse,draw,inner sep=-2pt,text width=2cm},
  -,shorten >= 0pt,shorten <= 0pt
]

\begin{scope}[start chain=going below,node distance=4mm]
\foreach \i in {1,2,...,4}
  \node[fsnode,on chain,minimum size=12pt,inner sep=0pt] (f\i) {\footnotesize{$a$}};
\node[double,fill=red!15,on chain,minimum size=12pt,inner sep=0pt] (f5) {\footnotesize{$b$}};
\end{scope}
  
\begin{scope}[xshift=2cm,yshift=0.0cm,start chain=going below,node distance=4mm]
\foreach \i in {1,2,...,4}
  \node[ssnode,on chain,minimum size=12pt,inner sep=0pt] (s\i) {\footnotesize{$a$}};
\node[double,fill=red!15,on chain,minimum size=12pt,inner sep=0pt] (s5) {\footnotesize{$b$}};
\end{scope}

\draw (f1) -- (s2);
\draw (f1) -- (s3);
\draw (f1) -- (s4);
\draw (f1) -- (s5);
\draw (f2) -- (s1);
\draw (f2) -- (s3);
\draw (f2) -- (s4);
\draw (f2) -- (s5);
\draw (f3) -- (s1);
\draw (f3) -- (s2);
\draw (f3) -- (s4);
\draw (f3) -- (s5);
\draw (f4) -- (s1);
\draw (f4) -- (s2);
\draw (f4) -- (s3);
\draw (f4) -- (s5);
\draw (s1) -- (f5);
\draw (s2) -- (f5);
\draw (s3) -- (f5);
\draw (s4) -- (f5);
\end{tikzpicture}

\caption{Complete graph with 5 vertices and 1 marked vertex  and its associated bipartite graph. The marked vertex is labeled as $b$ and the non-marked vertices as $a$.}
\label{fig:cg}
\end{figure}
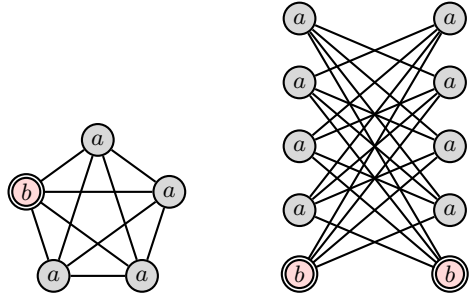

Due to the symmetry of the complete graph we can identify edges which will present the same behavior during the evolution of the quantum walk, that means they will have the same amplitudes. 
Some of the ideas used here are analogous from the ones used in the analysis of the complete graph in the coined model by Ref.~\cite{Wong:2015}.
For one marked vertex, label it as $b$ and the other non-marked vertices as $a$, see Fig.~\ref{fig:cg}. We can see that there are three types of edges: $\{a,a\}, \{a,b\}$ and $\{b,a\}$.  
Therefore, the set of the following vectors 
\begin{eqnarray}
\ket{a,a} &\coloneqq& \frac{1}{\sqrt{(n-1)(n-2)}}\sum_{\begin{subarray}{c}
x, y \in X\backslash M\\
 x\neq y
\end{subarray}}\ket{x,y},\\
\ket{a,b} &\coloneqq& \frac{1}{\sqrt{n-1}}\sum_{x\in X\backslash M, y\in M}\ket{x,y},\\
\ket{b,a} &\coloneqq& \frac{1}{\sqrt{n-1}}\sum_{x\in M, y\in X\backslash M}\ket{x,y},
\end{eqnarray}
forms an invariant subspace where the quantum walk takes place, that is, if we apply the evolution operator $U_M$ to any linear combination of these vectors, the result will be another linear combination of them.

By applying the evolution operator to the vectors in the invariant subspace and using that
\begin{equation}
U_P\ket{x,y} = 4\sqrt{p_{xy}}\sum_{y'\in X}\sqrt{p_{xy'}p_{y'x}}\ket{\psi_{y'}}-2\sqrt{p_{xy}}\ket{\psi_y}-2\sqrt{p_{xy}}\ket{\phi_x}+\ket{x,y},
\end{equation}
we obtain the reduced operator $U_r$
\begin{equation}
U_r = \left[\begin{array}{ccc}\cos^2\phi & \cos\phi\sin\phi & -\sin\phi \\\sin\phi & -\cos\phi & 0 \\\cos\phi\sin\phi & \sin^2\phi & \cos\phi\end{array}\right],
\end{equation}
where
\[\cos\phi = \frac{n-3}{n-1},\textrm{\quad and \quad}\sin\phi = \frac{2\sqrt{n-2}}{n-1}.\]
Thanks to the invariant subspace, we reduced our problem to a $3$-dimensional one. Now it becomes easier to find the spectrum of the reduced operator, so we can have a complete description of the system. The orthonormalized eigenvectors of $U_r$ associated with eigenvalues $\lambda = -1, e^{i\theta}, e^{-i\theta}$ are, respectively,
\begin{eqnarray}
\ket{v_{-1}} &=& \frac{1}{\sqrt{n^2-3n+3}}\left[\begin{array}{c}\sqrt{n-2} \\2-n \\1\end{array}\right],\\
\ket{v_{+}} &=& \frac{\sqrt{(n-2)(n-1)}}{\sqrt{2(n^2-3n+3)}}\left[\begin{array}{c}\frac{-1+i\sqrt{(n-2)(n^2-3n+3)}}{(n-1)\sqrt{n-2}} \\\frac{n-1}{n-2-i\sqrt{(n-2)(n^2-3n+3)}} \\1\end{array}\right],\\
\ket{v_{-}} &=& \ket{v_{+}}^*,
\end{eqnarray}
where $\ket{\cdot}^*$ stands for the complex conjugate of $\ket{\cdot}$, 
\[\cos\theta = \frac{1+\cos^2\phi}{2},\textrm{\quad and \quad}\sin\theta = \frac{\sin\phi\sqrt{4-\sin^2\phi}}{2}.\]

\begin{figure}[!htb]
\centering
\includegraphics[width=2.5in]{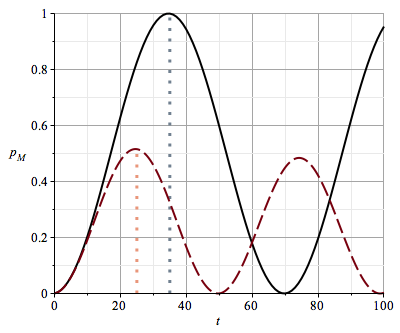}
\caption{Probability of finding a marked vertex on the complete graph with $n=2000$ and $m=1$ for 100 steps of the walk. For $U_M$ (solid line), the maximum probability is approximately $1$ at $t = 35$. For $U_{P'}$ (dashed line),  the maximum probability is approximately $0.52$ at $t = 25$.} \label{fig:m1}
\end{figure}

The behavior of $U_M$ is depicted in Fig.~\ref{fig:m1}. We can see that the probability of obtaining a marked vertex after measurement goes to 1 which is higher than the maximum probability for the evolution using $U_{P'}$, as described in Sec.~\ref{sec:sw}.
Our goal is to prove that. The main idea is to show that the initial state is roughly $\ket{a,a}$ and from there, by applying the evolution operator a certain number of times, we can get to the state $\ket{b,a}$, which is our target state. Note that if we measure the system which is in the state $\ket{b,a}$, it will return us a marked vertex with probability 1.

The initial state $\ket{\psi(0)}$, given by Eq.~(\ref{ht_ini_cond_CG}), is a uniform superposition over all edges of the graph. We can express it as
\begin{equation}
\begin{split}
\ket{\psi(0)} &= \frac{1}{\sqrt{n}}\left(\sqrt{n-2}\ket{a,a}+\ket{a,b}+\ket{b,a}\right).
\end{split}
\end{equation}
For large $n$, $\ket{\psi(0)} \approx \ket{a,a}$. And $\ket{a,a}$ can be expressed as a combination of the eigenvectors $\ket{v_+}$ and $\ket{v_-}$, that is,
\begin{equation}
\ket{a,a} = \left[\begin{array}{c}1 \\0 \\0\end{array}\right] \approx \frac{i}{\sqrt{2}}\left(-\ket{v_+}+\ket{v_-}\right) = \left[\begin{array}{c}{\frac{\sqrt{n-1}}{\sqrt{n-2}}} \\\frac{1}{\sqrt{n-1}} \\0\end{array}\right].
\end{equation}
Then, the state of the system at time $t$ is
\begin{equation}
\ket{\psi(t)} = U_r^t\ket{\psi(0)} \approx \frac{i}{\sqrt{2}}\left(-U_r^t\ket{v_+}+U_r^t\ket{v_-} \right) = \frac{i}{\sqrt{2}}\left(-e^{i\theta t}\ket{v_+}+e^{-i\theta t}\ket{v_-}\right).
\end{equation}
Our goal is to obtain the target state $\ket{b,a}$. By doing
\begin{equation}\label{eq:tf1}
t_f = \frac{\pi}{2\theta}  \approx \frac{\pi}{4}\sqrt{n},
\end{equation}
we get that our final state will be roughly $\ket{b,a}$, that is,
\begin{equation}
\ket{\psi(t_f)} \approx \frac{1}{\sqrt{2}}\left(\ket{v_+}+\ket{v_-}\right) = \left[\begin{array}{c}-\frac{1}{\sqrt{n^2-3n+3}\sqrt{n-1}} \\ \frac{\sqrt{n-2}}{\sqrt{n^2-3n+3}\sqrt{n-1}} \\\frac{\sqrt{n-2}\sqrt{n-1}}{\sqrt{n^2-3n+3}}\end{array}\right] \approx \left[\begin{array}{c}0 \\0 \\1\end{array}\right] = \ket{b,a}. 
\end{equation}

Note that the final complexity of the algorithm will be $O(\sqrt{n})$, according to Eq.~(\ref{eq:tf1}). The same as in the algorithm using $U_{P'}$, see Sec.~\ref{sec:sw} for comparison.

\subsection{Multiple marked vertices}

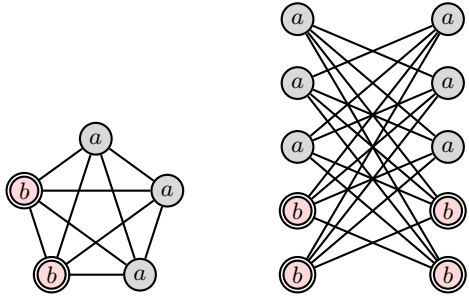
\begin{figure}[!htb]
\centering
\begin{tikzpicture}[thick,shorten >=0pt,-]
  \tikzstyle{vertex}=[circle,draw=black,fill=black!15,minimum size=12pt,inner sep=0pt]
\tikzstyle{vertexm}=[double,circle,draw=black,fill=red!15,minimum size=12pt,inner sep=0pt]
  \foreach \name/\angle/\text in { 
                                  P-1/90/a, P-2/18/a, P-3/-54/a}
    \node[vertex,xshift=6cm,yshift=.5cm] (\name) at (\angle:1cm) {\footnotesize{$\text$}};

\node[vertexm,xshift=6cm,yshift=.5cm] (P-4) at (234:1cm) {\footnotesize{$b$}};
\node[vertexm,xshift=6cm,yshift=.5cm] (P-5) at (162:1cm) {\footnotesize{$b$}};
  \foreach \from/\to in {1/2,2/3,3/4,4/5,5/1,1/3,2/4,3/5,4/1,5/2}
    { \draw (P-\from) -- (P-\to);}

\end{tikzpicture}
\hspace{1cm}
\begin{tikzpicture}[thick,
  every node/.style={draw,circle},
  fsnode/.style={fill=black!15},
  ssnode/.style={fill=black!15},
  every fit/.style={ellipse,draw,inner sep=-2pt,text width=2cm},
  -,shorten >= 0pt,shorten <= 0pt
]

\begin{scope}[start chain=going below,node distance=4mm]
\foreach \i in {1,2,...,3}
  \node[fsnode,on chain,minimum size=12pt,inner sep=0pt] (f\i) {\footnotesize{$a$}};
\node[double,fill=red!15,on chain,minimum size=12pt,inner sep=0pt] (f4) {\footnotesize{$b$}};
\node[double,fill=red!15,on chain,minimum size=12pt,inner sep=0pt] (f5) {\footnotesize{$b$}};
\end{scope}
  
\begin{scope}[xshift=2cm,yshift=0.0cm,start chain=going below,node distance=4mm]
\foreach \i in {1,2,...,3}
  \node[ssnode,on chain,minimum size=12pt,inner sep=0pt] (s\i) {\footnotesize{$a$}};
\node[double,fill=red!15,on chain,minimum size=12pt,inner sep=0pt] (s4) {\footnotesize{$b$}};
\node[double,fill=red!15,on chain,minimum size=12pt,inner sep=0pt] (s5) {\footnotesize{$b$}};
\end{scope}

\draw (f1) -- (s2);
\draw (f1) -- (s3);
\draw (f1) -- (s4);
\draw (f1) -- (s5);
\draw (f2) -- (s1);
\draw (f2) -- (s3);
\draw (f2) -- (s4);
\draw (f2) -- (s5);
\draw (f3) -- (s1);
\draw (f3) -- (s2);
\draw (f3) -- (s4);
\draw (f3) -- (s5);
\draw (f4) -- (s1);
\draw (f4) -- (s2);
\draw (f4) -- (s3);
\draw (f4) -- (s5);
\draw (s1) -- (f5);
\draw (s2) -- (f5);
\draw (s3) -- (f5);
\draw (s4) -- (f5);
\end{tikzpicture}

\caption{Complete graph with 5 vertices and 2 marked vertices and its associated bipartite graph. The marked vertices are labeled as $b$ and the non-marked vertices as $a$.}
\label{fig:cgm}
\end{figure}

The analysis for multiple marked vertices is similar to the analysis with one marked vertex. The difference, in this case, is that multiple marked vertices imply we have edges between two marked vertices, see Fig.~\ref{fig:cgm} for instance. Therefore, there are four types of edges $\{a,a\}, \{a,b\}, \{b,a\}$ and $\{b,b\}$ and we end up dealing with a 4-dimensional problem.
Our invariant subspace is now described by the following vectors:
\begin{eqnarray}
\ket{a,a} &\coloneqq& \frac{1}{\sqrt{(n-m)(n-m-1)}}\sum_{\begin{subarray}{c}
x, y \in X\backslash M\\
 x\neq y
\end{subarray}}\ket{x,y}\\
\ket{a,b} &\coloneqq& \frac{1}{\sqrt{m(n-m)}}\sum_{x\in X\backslash M, y\in M}\ket{x,y}\\
\ket{b,a} &\coloneqq& \frac{1}{\sqrt{m(n-m)}}\sum_{x\in M, y\in X\backslash M}\ket{x,y}\\
\ket{b,b} &\coloneqq& \frac{1}{\sqrt{m(m-1)}}\sum_{\begin{subarray}{c}
x, y \in M\\
 x\neq y
\end{subarray}}\ket{x,y}
\end{eqnarray}
Note that states $\ket{a,a}, \ket{a,b}, \ket{b,a}$ are the same as in the one marked vertex case, by substituting $m=1$.
The reduced operator is given by
\begin{equation}
U_r = \left[\begin{array}{cccc}\cos^2\phi & \cos\phi\sin\phi & -\cos\theta\sin\phi & -\sin\theta\sin\phi \\\cos\theta\sin\phi & -\cos\theta\cos\phi & -\sin^2\theta & \cos\theta\sin\theta \\\cos\phi\sin\phi & \sin^2\phi & \cos\theta\cos\phi & \sin\theta\cos\phi\\ \sin\theta\sin\phi & -\sin\theta\cos\phi & \cos\theta\sin\theta & -\cos^2\phi\end{array}\right],
\end{equation}
where
\[\cos\phi = \frac{n-2m-1}{n-1},\textrm{\quad and \quad}\sin\phi = \frac{2\sqrt{m(n-m-1)}}{n-1};\]
\[\cos\theta = \frac{n-2m+1}{n-1},\textrm{\quad and \quad}\sin\theta = \frac{2\sqrt{(n-m)(m-1)}}{n-1}.\]

\begin{figure}[!htb]
\centering
\includegraphics[width=2.5in]{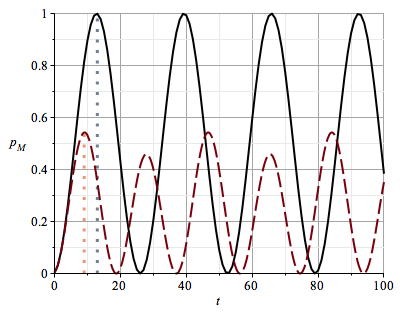}
\caption{Probability of finding a marked vertex on the complete graph with $n=2000$ and $m=7$ for 100 steps of the walk. For $U_M$ (solid line), the maximum probability is approximately $1$ at $t = 13$. For $U_{P'}$ (dashed line),  the maximum probability is approximately $0.54$ at $t = 9$.} 
\label{fig:m2}
\end{figure}

The behavior for multiple marked vertices is the same as for the one marked vertex case. Fig.~\ref{fig:m2} shows that the probability of obtaining a marked vertex goes to 1 differently from what happens to the evolution using $U_{P'}$, as described in Sec.~\ref{sec:sw}.

In order to analyze this case, we need to find some approximations. 
Note that $\cos \theta \approx \cos \phi$ for large $n$. By doing $\cos\theta = \cos\phi$, two of the eigenvalues of $U_r$ reduce to $e^{\pm i\phi}$ associated to the following approximated eigenvectors
\begin{equation}
\ket{v_+} = \frac{\sqrt{1-\cos\phi}}{2}\left[\begin{array}{c}\frac{\sin^2\phi\cos\phi[(3-2\cos\phi)\cos\phi-i\sin\phi(3-\cos\phi)]}{(1-\cos\phi)^2[\cos\phi(3+2\cos\phi-\cos^2\phi)-i\sin\phi(1+\cos\phi-2\cos^2\phi)]} \\\frac{2\cos\phi+i\sin\phi}{\sin\phi-2i\cos\phi} \\\frac{\sin\phi}{1-\cos\phi} \\1 \end{array}\right],
\end{equation}
and $\ket{v_-} = \ket{v_+}^*$, which are going to be useful as we can express the initial state by a linear combination of them.

The initial state written in the invariant basis is
\begin{equation}
\begin{split}
\ket{\psi(0)} &= \frac{1}{\sqrt{n(n-1)}}\left(\sqrt{(n-m)(n-m-1)}\ket{a,a}+\sqrt{m(n-m)}\ket{a,b}+\right.\\
&\left.+\sqrt{m(n-m)}\ket{b,a}+\sqrt{m(m-1)}\ket{b,b}\right).
\end{split}
\end{equation}
Again, $\ket{\psi(0)}$ is roughly $\ket{a,a}$  and 
\begin{equation}
\ket{a,a} = \left[\begin{array}{c}1 \\0 \\0 \\0\end{array}\right] \approx \frac{i}{\sqrt{2}}\left(-\ket{v_+}+\ket{v_-}\right).
\end{equation}
The state of the system at time $t$ is
\begin{equation}
\ket{\psi(t)} = U_r^t\ket{\psi(0)} \approx \frac{i}{\sqrt{2}}\left(-e^{i\phi t}\ket{v_+}+e^{-i\phi t}\ket{v_-}\right).
\end{equation}
At time
\begin{equation}
t_f = \frac{\pi}{2\phi}  \approx \frac{\pi}{4}\sqrt{\frac{n}{m}},
\end{equation}
our final state will be roughly $\ket{b,a}$, that is,
\begin{equation}
\ket{\psi(t_f)} \approx \frac{1}{\sqrt{2}}\left(\ket{v_+}+\ket{v_-}\right) = \left[\begin{array}{c}0 \\0 \\\frac{\sqrt{n-m-1}}{\sqrt{n-1}} \\\frac{\sqrt{m}}{\sqrt{n-1}}\end{array}\right] \approx \left[\begin{array}{c}0 \\0 \\1 \\0\end{array}\right] = \ket{b,a}. 
\end{equation}

\section{Other search operators}\label{sec:oso}

Recently, Portugal {\it et al.}~\cite{Portugal:2015} have shown that Szegedy's quantum walk is contained into the staggered quantum walk model. 
The latter model also uses queries for searching. Originally, Falk~\cite{Falk:2013} used the operator ${\cal R}_M{\cal R}_{\cal B}{\cal R}_M{\cal R}_{\cal A}$ to search in the two-dimensional lattice, which was later shown by \cite{Portugal:2015} that it was in fact the non-planar regular graph of degree $6$. Inspired by Falk~\cite{Falk:2013}, we numerically analyze the behavior of different operators, by making different combinations with the reflections around the marked vertices. The following operators will be compared:
\begin{eqnarray}
U_1 &= &U_M = {\cal R}_{\cal B}{\cal R}_{\cal A}{{\cal R}_M}_1,\\
U_2 &= &U_{P'} = {\cal R}_{{\cal B}'}{\cal R}_{{\cal A}'},\\
U_3 &= &{\cal R}_{\cal B}{{\cal R}_M}_2{\cal R}_{\cal A}{{\cal R}_M}_1,\\
U_4 &= &{\cal R}_{\cal B}{{\cal R}_M}_1{\cal R}_{\cal A}{{\cal R}_M}_1,\\
U_5 &= &{{\cal R}_M}_1{\cal R}_{\cal B}{{\cal R}_M}_1{\cal R}_{\cal A}.
\end{eqnarray}

\begin{figure}[!htb]
\centering
\subfigure[$m=1$]{\label{fig:torusa}
\includegraphics[width=2.5in]{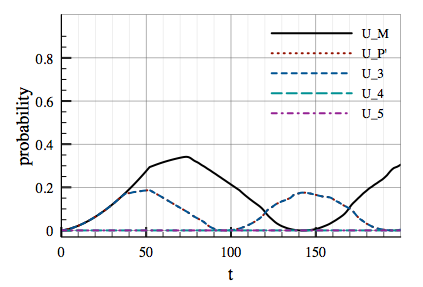}}
\subfigure[$m=2$]{\label{fig:torusb}
\includegraphics[width=2.5in]{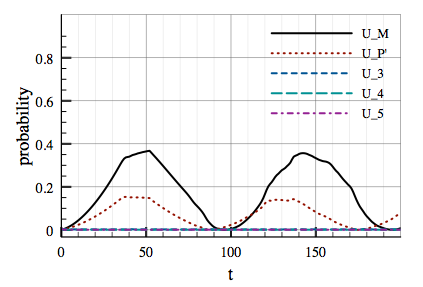}}

\caption{Probability of finding a marked vertex on the two-dimensional lattice with $53\times 53$ vertices for 200 steps.  (a) With one marked vertex ($m=1$). (b) With two marked vertices ($m=2$). } \label{fig:torus}
\end{figure}

Fig.~\ref{fig:torus} shows the probability $p_M$ for these evolution operators for one and two marked vertices in the two-dimensional lattice with torus-like boundary conditions.
We decided to apply these operators to other graph than the complete graph, to show that the operator $U_M$ can achieve higher probability than the operator $U_{P'}$ for other graphs. However, for the two-dimensional lattice, $U_M$ and $U_{P'}$ will act the same if the number of vertices is even, the difference between them only occur when $n$ is odd, as we can clearly see in Fig.~\ref{fig:torus}.

The value of $p_M$ is constant for $U_4$. It is easy to show that, for any graph, ${{\cal R}_M}_1{\cal R}_{\cal A}{{\cal R}_M}_1 = {\cal R}_{\cal A}$. Therefore, $U_4 = {\cal R}_{\cal B}{\cal R}_{\cal A} = U_P$ and $\ket{\psi(0)}$ is its $1$-eigenvector. Differently from $U_4$, the behavior of $U_5$ varies a little bit and its probability stays very close to the initial one, which we can only identify by looking at the resulting data as the scale in the figure does not allow us to see it.

Moreover, we can see that there is a superposition of the curves for $U_3$ and $U_{P'}$ in Fig~\ref{fig:torusa}. They will have the same behavior, independent on the parity of $n$, but dependent on the value of $m$. If $m > 1$ the curves will be different, as shown in Fig.~\ref{fig:torusb}. Those operators are not the same when $m=1$ but their action on the initial state is. We will prove below that this is valid for some class of graphs.

\subsection{Equivalence between $U_3$ and $U_{P'}$}

For any strongly regular graph{\footnote{A strongly regular graph is a regular graph where every two adjacent vertices have $\lambda$ common neighbors and every two non-adjacent vertices have $\mu$ common neighbors.}} with only one marked vertex ($m=1$), the action of $U_3$ on $\ket{\psi(0)}$ is the same as $U_{P'}$, that is,
\begin{equation}
U_3^t\ket{\psi(0)} = ({\cal R}_{\cal B} {{\cal R}_M}_2 {\cal R}_{\cal A} {{\cal R}_M}_1)^t\ket{\psi(0)} = U_{P'}^t\ket{\psi(0)} = (R_{{\cal B}'}R_{{\cal A}'})^t\ket{\psi(0)}\quad \forall t.
\end{equation}

The idea is to show that the application of $U_{P'}^t$ to $\ket{\psi(0)}$ will be a combination of the states
$$
\ket{\phi_1} = \sum_{x\neq b}\ket{\Phi_x},\quad \ket{\phi_2} = \sum_{x\neq b}p_{xb}\ket{\Phi_x},
$$
$$
\ket{\psi_1} = \sum_{y\neq b}\ket{\Psi_y},\quad \ket{\psi_2} = \sum_{y\neq b}p_{yb}\ket{\Psi_y},{\textrm{ and }}\ket{\Phi_b}.
$$
where $b$ is the marked vertex. And then show that $U_{P'}-U_3$ applied to each one of this states will give us 0.
Note that 
\begin{equation}
\ket{\psi(0)} = \frac{1}{\sqrt{n}}\ket{\phi_1}+\frac{1}{\sqrt{n}}\ket{\Phi_b}.
\end{equation}

By using that $P$ is symmetric and stochastic (its rows add up to one), and from the fact that two adjacent vertices have $\lambda$ common neighbors and two non-adjacent vertices have $\mu$ common neighbors, we obtain 
\begin{eqnarray}
R_{{\cal A}'}\ket{\phi_1} &=& \ket{\phi_1},\\
R_{{\cal A}'}\ket{\phi_2} &=& \ket{\phi_2},\\
R_{{\cal A}'}\ket{\psi_1} &=& 2\ket{\phi_1} -2\ket{\phi_2}-\ket{\psi_1},\\
R_{{\cal A}'}\ket{\psi_2} &=& \frac{2\mu}{k^2}\ket{\phi_1} +\frac{2(\lambda-\mu)}{k}\ket{\phi_2}-\ket{\psi_2}, \\
R_{{\cal A}'}\ket{\Phi_b} &=& -\ket{\Phi_b},
\end{eqnarray}
where $k$ is the degree of the graph. Analogously,
\begin{eqnarray}
R_{{\cal B}'}\ket{\phi_1} &=& 2\ket{\psi_1} -2\ket{\psi_2}-\ket{\phi_1},\\
R_{{\cal B}'}\ket{\phi_2} &=& \frac{2\mu}{k^2}\ket{\psi_1} +\frac{2(\lambda-\mu)}{k}\ket{\psi_2}-\ket{\phi_2}, \\
R_{{\cal B}'}\ket{\psi_1} &=& \ket{\psi_1},\\
R_{{\cal B}'}\ket{\psi_2} &=& \ket{\psi_2},\\
R_{{\cal B}'}\ket{\Phi_b} &=& 2\ket{\psi_2}-\ket{\Phi_b}.
\end{eqnarray}
Therefore, we can conclude that 
\begin{equation}
U_{P'}^t\ket{\psi(0)} = \alpha_1(t)\ket{\phi_1}+\alpha_2(t)\ket{\phi_2}+\alpha_3(t)\ket{\psi_1}+\alpha_4(t)\ket{\psi_2}+\alpha_5(t)\ket{\Phi_b},
\end{equation}
which means that any number of applications of the operator $U_{P'}$ to the initial state $\ket{\psi(0)}$ will be a linear combination of the states $\ket{\phi_1}, \ket{\phi_2}, \ket{\psi_1}, \ket{\psi_2}$ and $\ket{\Phi_b}$. 
Moreover, $R_{{\cal A}'}U_{P'}^t\ket{\psi(0)}$ is also a linear combination of those states.
From this fact, we are left to show that $R_{\cal A} R_{M_1} - R_{{\cal A}'}$ applied to any of these states is 0 and that $R_{\cal B} R_{M_2} - R_{{\cal B}'}$ applied to any of these states is 0. Since the two latter claims are symmetric, it suffices to show for $R_{\cal A} R_{M_1} - R_{{\cal A}'}$.

Note that $R_{\cal A}R_{M_1} \ket{\Phi_x} = R_{{\cal A}'}\ket{\Phi_x} = \ket{\Phi_x}$, for all $x\neq b$. Then, the claim is true for $\ket{\phi_1}$ and $\ket{\phi_2}$. For $y\neq b$, we have,
\begin{equation}\label{eq:psiy1}
(R_{\cal A}R_{M_1}-R_{{\cal A}'})\ket{\Psi_y} = -2p_{yb}\ket{\Phi_b}+2\sqrt{p_{yb}}\ket{b,y},
\end{equation}
which implies that the claim is also true for $\ket{\psi_1}$ and $\ket{\psi_2}$.
And, finally, we have $R_{\cal A}R_{M_1}\ket{\Phi_b} = R_{{\cal A}'}\ket{\Phi_b} = -\ket{\Phi_b}$. Therefore $U_3$ is equivalent to $U_{P'}$ when applied to the initial state $\ket{\psi(0)}$.

\section{Conclusions and Discussions}\label{sec:conc}

Queries are usually used to differentiate marked vertices when searching with quantum walks. Here we analyzed, for the first time, the behaviour of using standard queries when searching with Szegedy's quantum walk model. Instead of using the usual way of searching, by the operator $U_{P'}$ obtained from the absorbing walk, we take the usual evolution operator, $U_P$ (without marked vertices), and we add a reflection around the marked vertices acting in the first register. In this way, we can change Szegedy's way of searching and boost the success probability to 1 for finding a marked vertex on the complete graph. The analysis is done for one and multiple marked vertices. Note that the complexity of the algorithm remains the same. 


Numerical simulations in the two-dimensional lattice, with torus-like boundary conditions, showed that the new operator can achieve higher probability than $U_{P'}$ for other graphs than the complete graph. We also tried to use more than one reflection around the marked vertices, similar to the search operator by Falk~\cite{Falk:2013} for the staggered model, and that didn't make any improvement.
Recently, Wong~\cite{Wong:2015} showed a similar result, boosting the probability to 1 for the coined model by adding self-loops to each vertex in the complete graph. And Pr\={u}sis~{\it et al.}~\cite{Prusis:2015} showed how to double the success probability for searching in the coined quantum walk model by using internal state measurements. 
Adding self-loops in Szegedy's model, which means making the walk lazy, doesn't seem to improve searching when using $U_{P'}$ or $U_M$, according to some numerical results.

We showed that the operator $U_{P'}$ can be written using a combination of reflections around the marked vertex acting in the first and second register for strongly regular graphs with only one marked vertex.
For the operator $U_{P'}$, we have a partial description of its eigenvalues and eigenvectors obtained from Szegedy's spectral theorem~\cite{Szegedy:2004}. Unfortunately, we can't use the same description for $U_M$ as it is not a product of two reflections. Therefore we should find its eigenvalues and eigenvectors using other methods. 
Note that all operators described in this paper can be cast into the staggered quantum walk model, since the reflections ${{\cal R}_M}_1$ and ${{\cal R}_M}_2$ are partial orthogonal reflections, see~\cite{Portugal:2015,Portugal:2016}.




\section*{Acknowledgments}
The author thanks R. Portugal and A. Ambainis for useful comments.
This work was supported by the EU FP7 project QALGO (Grant Agreement No. ESRTD-7IP-13).

\bibliographystyle{abbrv}
\bibliography{mybib}  

\end{document}